\documentclass[%
 reprint,
superscriptaddress,
 amsmath,amssymb,
 aps,
]{revtex4-2}
\usepackage{graphicx}
\usepackage{dcolumn}
\usepackage{bm}
\usepackage{mathtools}%
\usepackage{placeins}
\usepackage{float}
\usepackage[caption = false]{subfig}
\usepackage{amsmath,array}
\usepackage[table]{xcolor}
\PassOptionsToPackage{breaklinks}{hyperref}
\usepackage{xurl} 
\usepackage[colorlinks,linkcolor=blue,filecolor=blue,urlcolor=blue,citecolor=blue]{hyperref}

\DeclarePairedDelimiterX\braket[2]{\langle}{\rangle}{#1 \delimsize\vert #2}

\begin{document}

\preprint{APS/123-QED}

\title{Single-shot error correction on toric codes with high-weight stabilizers}

\author{Yingjia Lin}%
\email{yingjia.lin@duke.edu}
\affiliation{Duke Quantum Center, Duke University, Durham, NC 27701, USA}
\affiliation{
Department of Physics, Duke University, Durham, NC 27708, USA
}
\author{Shilin Huang}%
\altaffiliation{Current Address: Department of Applied Physics, Yale University}
\email{shilin.huang@yale.edu}
\affiliation{Duke Quantum Center, Duke University, Durham, NC 27701, USA}
\affiliation{Department of Electrical and Computer Engineering, Duke University, Durham, NC 27708, USA}

\author{Kenneth R. Brown}%
\email{ken.brown@duke.edu}
\affiliation{Duke Quantum Center, Duke University, Durham, NC 27701, USA}
\affiliation{
Department of Physics, Duke University, Durham, NC 27708, USA
}
\affiliation{Department of Electrical and Computer Engineering, Duke University, Durham, NC 27708, USA}
\affiliation{Department of Chemistry, Duke University, Durham, NC 27708, USA}

\date{\today}

\begin{abstract}
For quantum error correction codes the required number of measurement rounds typically increases with the code distance when measurements are faulty. Single-shot error correction allows for an error threshold with only one round of noisy syndrome measurements regardless of the code size. Here we implement single-shot check operators for toric codes. The single-shot checks are constructed by Gaussian elimination following Campbell \cite{campbell2019theory}. The single-shot check operators result in a sustainable threshold at 5.62\% for an error model with noisy measurements, outperforming the conventional toric code check operators with multiple rounds of noisy measurement. 
The cost of the transformation is non-local high-weight stabilizer generators. 
We then consider a gate-based error model that leads to increased measurement error with stabilizer weight. Here we find no single-shot threshold behavior and instead find the code family will have an optimal code size for a fixed error rate. For this error model, the conventional check operators with multiple measurements yields a lower logical error rate.
\end{abstract}

\maketitle

Noise leads to a loss of quantum information, and poses a challenge for quantum computers. To overcome this problem, quantum error correction codes are proposed where check operators are measured to detect and correct errors. In practice, the measurement results of check operators are not reliable [\onlinecite{chen2021exponential}]. A correction based on these faulty syndrome might cause further accumulation of errors. A common way to overcome this problem is to repeat the measurements \cite{548464,wang2003confinement,RaussendorfPRL2007,PhysRevA.80.052312}. However, such methods require additional time and resources. As an alternative solution, single-shot error correction is proposed where each check operator is measured only once [\onlinecite{PhysRevX.5.031043}]. Some examples of codes that support single-shot error corrections include 3D gauge color codes \cite{PhysRevX.5.031043, bombin2015gauge,brown2016fault}, 3D subsystem toric codes \cite{kubica2022single},  4D toric codes \cite{dennis2002topological,campbell2019theory,breuckmann2016local}, hypergraph product codes \cite{campbell2019theory,higgott2023improved,zeng2019higher,10.1145/3434163}, quantum Tanner codes \cite{gu2023single} and hyperbolic codes \cite{breuckmann2021single}. Decoders have been developed for single-shot error correction that achieve comparable or even higher threshold than the threshold of the 2D toric code obtained from repeated measurements under specific noise models \cite{PRXQuantum.2.020340,brown2016fault,higgott2023improved,Grospellier_2021,kubica2022single,vasmer2021cellular}. Advantages of the single-shot error correction includes reduced time costs and resilience to time-correlated noise \cite{bombin2016resilience}. It has been noticed that the single-shot property is closely related to macroscopic energy barrier and self-correcting memories \cite{campbell2019theory,PhysRevX.5.031043,alicki2010thermal,roberts2020symmetry}.

Despite the advantages of single-shot error correction, not all code implementations are compatible with it [\onlinecite{campbell2019theory},\onlinecite{PRXQuantum.2.020340}]. For example,  single-shot error correction is impossible with conventional local checks [\onlinecite{campbell2019theory}] used for two-dimensional toric codes [\onlinecite{kitaev2003fault}]. However, any code can have achieve single-shot error correction if the generators of the stabilizer group to be measured are selected carefully \cite{campbell2019theory,delfosse2021beyond}. Based on this theory, we investigate a method to construct single-shot error correction generator checks for two-dimensional toric codes. The same method can be generalized to all the Calderbank-Shor-Steane (CSS) codes [\onlinecite{calderbank1996good}, \onlinecite{steane1996multiple}]. Our simulations show that the generators we choose exhibit a sustainable threshold at 5.62\% for the toric code for a phenomenological noise model, which includes data and measurement errors. This threshold is much higher than the toric code threshold at 2.9\% for the same noise model with local generator checks with multiple rounds of measurement [\onlinecite{wang2003confinement}]. A limitation in implementing this method is the need to measure non-local, high-weight generator checks and we investigate a gate-error model that shows the challenges of applying these large error checks in practice.

We begin with definitions of stabilizer codes and CSS codes. A stabilizer code is defined by a stabilizer group. A stabilizer group $S$ is an Abelian group of Pauli operators on $n$ qubits which does not include $-I$. The stabilizer code defined by the stabilizer group is the $+1$-eigenspace of all the elements in the stabilizer group. The error correction of the stabilizer code is typically performed by measuring a generator set of the stabilizer group as the check operators to obtain the syndrome and then correcting based on the syndrome. If there are $n-k$ operators in the generator set, the code encodes $k$ qubits. 

CSS codes \cite{calderbank1996good,steane1996multiple} have a generator set where each generator is composed of either Pauli $X$ or $Z$ operators, which we refer to as $X$-type checks or $Z$-type checks. $X$-type checks can only detect Pauli $Z$ errors, while $Z$-type checks can only detect Pauli $X$ errors. 
We can then write the $X(Z)$-type checks into an $X(Z)$ parity check matrix $H_{X(Z)}$. Each row of $H_{X(Z)}$ represents an $X(Z)$-type check. The matrix element $H_{X(Z)}^{ij}=1$ when the $j$th qubit is acted on by an $X(Z)$ operator from the $i$th $X(Z)$-type check. The condition $H_ZH_X^T=0$ is imposed so that all the checks commute with each other. In these matrices, the product of two $X(Z)$-type checks can be mapped to the binary addition of the corresponding rows of $H_X(Z)$. Therefore, if we perform reversible row transformations on $H_X$ and $H_Z$, we can obtain a new set of checks for the same code.

Single-shot error correction is a property of the parity check matrices. Bomb\'{i}n first observed that for certain code families, decoding with a faulty syndrome only leads to confined residual errors [\onlinecite{bombin2015gauge}]. For these codes, a single round of measurement is sufficient for error correction.

Single-shot error correction is typically performed in two steps: syndrome decoding and qubit decoding [\onlinecite{campbell2019theory}]. The syndrome decoding aims at correcting measurement errors. When the parity check matrix $H$ of the code is redundant, we can define a parity check matrix $H_m$ for the syndrome, which we call the metacheck parity matrix. If we obtain a syndrome $S$, a decoder can be exploited to find a syndrome correction $S_c$ such that $H_m(S+S_c)=0$. The syndrome decoding process benefits from redundant parity checks that encode the syndrome information into a classical code\cite{PhysRevA.90.062304,ashikhmin2020quantum,delfosse2021beyond, nemec2023quantum}.

Syndrome decoding is then followed by qubit decoding. In the qubit-decoding process, the decoder takes the corrected syndrome $S+S_c$ and applies a correction $E_c$ on qubits where $S+S_c=HE_c$. We can then use decoders assuming perfect syndrome measurement. 

When the parity check matrix is full rank, we can skip syndrome decoding as metachecks do not exist and we choose a correction based directly on the noisy syndrome. In practice, it is even possible to perform syndrome decoding and qubit decoding simultaneously \cite{Grospellier_2021,higgott2023improved}. 

After one round of single-shot error correction, errors that remain in the system are called residual errors. The weight of an error ${\rm wt}(E)$ is the number of qubits that are acted on non-trivially by the error chain $E$. We define the reduced weight of the residual error $E_R$ as ${\rm wt}_{r}(E_R)=\min\{{\rm wt}(sE_R)|s\in S\}$ where $S$ is the group of stabilizers of the codes. If the reduced weight of the residual error is bounded by a value that only depends on the number of measurement errors, this code is said to support single-shot error correction. 

Many code families do not support single-shot error correction with their conventional choices of check operators. However,  Campbell [\onlinecite{campbell2019theory}] showed that for any  stabilizer code there exists a choice of check operators that supports single-shot error correction without decreasing the effective code distance. Indeed, single-shot error correction is a property about the the choice of check operators rather than the code space itself. Precisely, we can always find a certain set of stabilizer generators $\{G_i\}$ as checks and a set of single-qubit error $\{E_i\}$ such that each generator $G_i$ anti-commutes with one distinct single-qubit error $E_i$. If a measurement error occurs on check $G_i$, a decoder can perceive it as a data-qubit error $E_i$. After error correction, the weight of the resulting residual errors introduced by the decoder is no more than the number of measurement errors.

We can construct single-shot generator checks for arbitrary CSS codes by using Gaussian elimination.  For convenience, we focus on correcting $X$ errors with $Z$ checks. By performing Gaussian elimination with reversible binary row transformations and relabeling of columns (qubits), the $Z$ parity-check matrix with $m$ rows and $n$ columns $H_Z$ can be transformed into a standard form $H_Z'=[I_{m\times m}, A_{m\times (n-m)}]$ when there are no redundant checks. In this representation, if a measurement error occurs when measuring a $Z$-type check $h_i$ representing by the $i$th row of $H_Z'$, the resulting syndrome is equivalent to the case when an additional $X$ error is added to the $i$th data qubit. A decoder that neglects the measurement error on $h_i$ will therefore introduce an $X$ error on the $i$th qubit after correction.
When the total number of errors is no more than half of the code distance, the reduced weight of the residual error is therefore no more than the number of measurement errors. 
We note that a similar idea has been exploited by Steane to avoid repetition in ancilla state verification \cite{steane2002fast}.


 In the toric code [\onlinecite{kitaev2003fault}] , two logical qubits are protected by $2L^2$ qubits residing on the edges of a $L\times L$ square lattice with periodic boundary conditions that forms a torus. For such a code block, the code distance is given by $L$. The toric code is a CSS code and the conventional choice generators yields two types of checks: the $X$-type stabilizers residing on the vertices, and the $Z$-type stabilizers residing on plaquettes. Both types are supported by the qubits surrounding them. We will refer to this conventional choice of checks as local checks.  
Since the $X$-type stabilizers and the $Z$-type stabilizers for toric codes are dual to each other, in the remaining parts of the paper, if not specified, we only discuss the $Z$-type stabilizers.

There are two types of single-shot $Z$-type checks that can be found by Gaussian elimination: 
\begin{itemize}
\item[(a)] Rectangular checks
$R_{i,j} = \prod_{k=i}^{L-1} \Box_{k,j}$, where $i = 0, \ldots, L-2$ and $j = 0, \ldots, L-1$.

\item[(b)] Circular checks $S_i = \prod_{m=i}^{L-2}\prod_{n=0}^{L-1} \Box_{n,m}$, where $i = 0, \ldots, L-2$. 
\end{itemize}
Here, the symbol $\Box_{i,j}$ refers to a local check defined on the plaquette at the location $(i,j)$. The circular checks are composed of plaquettes that have circled the torus.

\begin{figure}
\includegraphics[width=0.5\textwidth]{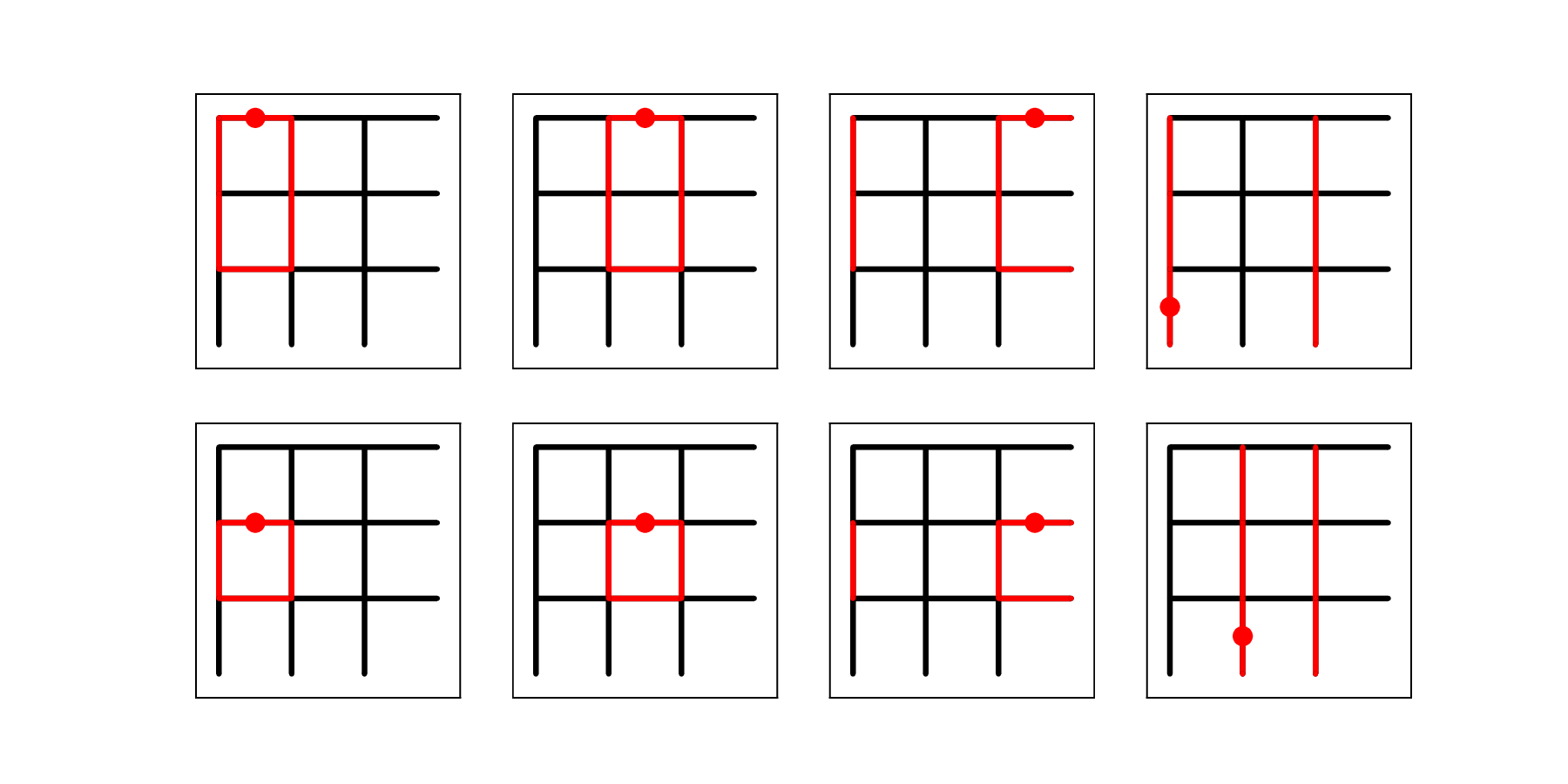}
\caption{\label{fig:3}The $Z$-type single-shot checks for $3\times 3$ toric codes. The eight checks are marked by red in the plots. The redundant check is already removed. The first six checks on the left are the rectangular checks. The last two checks on the right are the circular checks. We mark the corresponding single-qubit $X$ error, $E_C$,  with a red point on each graph. }
\end{figure}

An example of single-shot checks of a $3 \times 3$ toric code lattice is given in Fig.~\ref{fig:3}. The six graphs on the left show the rectangular stabilizers. The two graphs on the right show the circular stabilizers, which are rectangular checks that have circled the torus. Their corresponding single-qubit error positions are marked by red points in the graphs. 

For each check $R_{i,j}$, its corresponding single-qubit error $E_{R_{i,j}}$ acts on the horizontal qubit above plaquette $(i,j)$. For each $S_i$, its corresponding single-qubit error $E_{S_i}$ acts on the vertical qubit on the left of plaquette $(L-1,i)$. One can verify that each single-qubit error $E_C$ only flips its corresponding single-shot check $C$.

Quantum error correction noise models fall into three categories: code capacity, phenomenological, and circuit-level. Code capacity assumes perfect encoding, data errors, and then perfect syndrome measurement. In the code-capacity model, the choice of generators does not effect the threshold.

Single-shot error correction is usually studied in the phenomenological noise model. Errors are applied to the data and then the measurement associated with each check can be faulty.  Here we consider a single-sided error where we focus on correcting bit-flip $X$ errors with $Z$-type stabilizer measurements. The qubit error is an independent bit-flip channel $\mathcal{E} (\rho) = (1-p)\rho +pX\rho X$, where $p$ is the qubit error probability. Measurement errors are introduced by flipping each syndrome bit independently at the syndrome error probability $q$, where we set $q=p$. 

To evaluate the check performance, after $N$ rounds of single-shot error correction, we perform one extra round of error correction with qubit errors and noiseless syndromes following the technique in Ref. \cite{brown2016fault}.  If a logical error occurs after the perfect decoding, we consider it as a logical failure for the single-shot error correction and calculate the logical error rate $p_L$ for the single-shot error correction accordingly. In particular, we are interested in the threshold behavior. The threshold is the physical error rate under which we can achieve an arbitrarily low logical error rate $p_L$ by increasing the size of the lattice. We denote the threshold after $N$ rounds of single-shot error correction as $p_{\textrm{th}}(N)$. One can notice that $p_{\textrm{th}}(0)$ is exactly the toric code threshold for a noiseless syndrome. 

For qubit decoding, we use the minimum-weight-perfect-matching (MWPM) decoder \cite{edmonds_1965,edmonds1965maximum}, which has been successfully applied to various surface code models \cite{fowler2010surface,PhysRevX.11.031039,sundaresan2022matching,higgott2022fragile,roffe2023bias}. Specifically, we use the PyMatching package for simulations \cite{higgott2023sparse,higgott2022pymatching}. 

For the single-shot checks, we track the reversible row transformation we perform during the Gaussian elimination. After we obtain the syndrome with the single-shot checks, we first map the noisy syndrome back to syndrome of local checks by performing the inverse of these row transformations. We then pass the resulting syndrome to the toric code MWPM decoder to decode.

To confirm the single-shot property, we estimate the sustainable threshold $p_{\textrm{sus}}$ for the single-shot checks. The sustainable threshold was first introduced in studies of single-shot error correction on 3D gauge color codes \cite{brown2016fault}, and has been promoted to study the single-shot properties of other codes \cite{kubica2022single,higgott2023improved}. It is defined as the physical error rate $p_{\textrm{sus}}=\lim_{N\rightarrow \infty}p_{\textrm{th}}(N)$.

The existence of a non-zero $p_{\textrm{sus}}$ marks the single-shot property of the code. To obtain the sustainable threshold $p_{\textrm{sus}}$, we calculate $p_{\textrm{th}}(N)$ with $N$ ranging from 1 to 10 and fit the data with the following ansatz \cite{brown2016fault}:
\begin{equation}
    p_{\textrm{th}}(N)=p_{\textrm{sus}}\left(1-[1-p_{\textrm{th}}(0)/p_{\textrm{sus}}]e^{-\gamma N}\right)
\end{equation}

We determine the sustainable threshold $p_{\textrm{th}}$ for single-shot error correction to be 5.62\%. Compared with the phenomenological noise model threshold for toric codes at 2.9\% determined with repeated measurements [\onlinecite{wang2003confinement}], the single-shot checks provide a significant improvement on the code threshold. 

Notice that when $N=0$, $p_{\textrm{th}}(0)$ is $10.27\%$, which is the threshold for toric codes with an MWPM decoder and noiseless syndrome.  In Appendix \ref{sec:appenN1}, we calculate error rates for $N=1$ for local checks single-shot checks and observe no threshold for local checks. In Appendix \ref{sec:appen} we also calculate the sustainable threshold for both planar codes $5.51\%$ and rotated surface codes $5.54\%$ with our construction of single-shot checks.  

\begin{figure}

    \centering
    \includegraphics[width=0.45\textwidth]{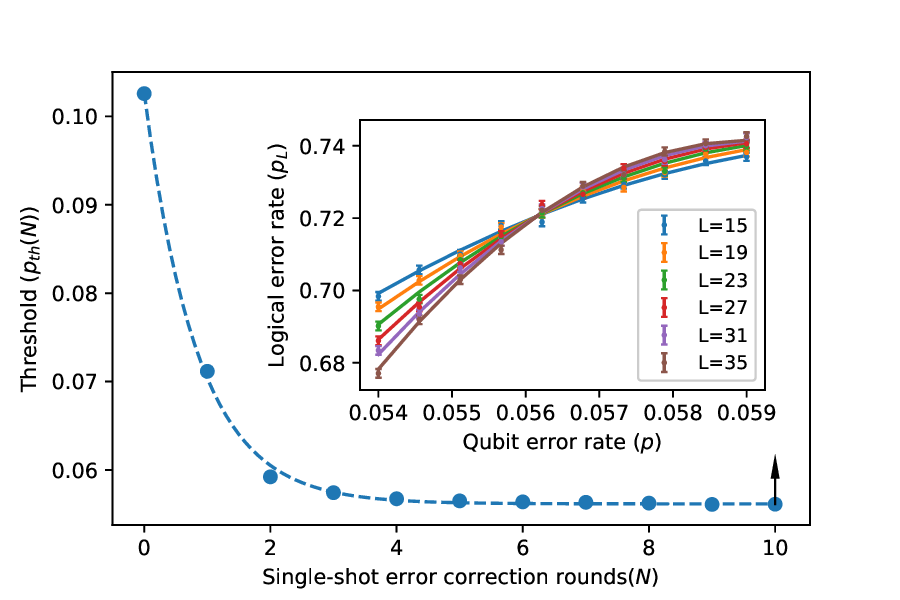}
    \caption{The threshold values scale with the rounds of single-shot error correction. Up to 10 rounds of single-shot error corrections are performed to estimate the sustainable threshold. The exponential function to fit for the threshold values is $p_{\textrm{th}}(N)=p_{\textrm{sus}}\{1-[1-p_{\textrm{th}}(0)/p_{\textrm{sus}}]e^{-\gamma N}\}$. The same equation has been used in \cite{brown2016fault}. The single-shot error correction sustainable threshold $p_{\textrm{sus}}$ is then determined at 5.62\%. The fitting parameters are determined at $\gamma=1.185$. $p_{\textrm{th}}(0)$ is fitted to be $0.1027$, which is the code capacity threshold for toric code with an MWPM decoder under bit-flip errors. The inset shows how we determine the threshold $p_{\textrm{th}}(N)$ for $N$ up to 10 by fitting the data around the threshold. }
    \label{fig:limit}
\end{figure}

\begin{figure}
    \centering
    \includegraphics[width=0.5\textwidth]{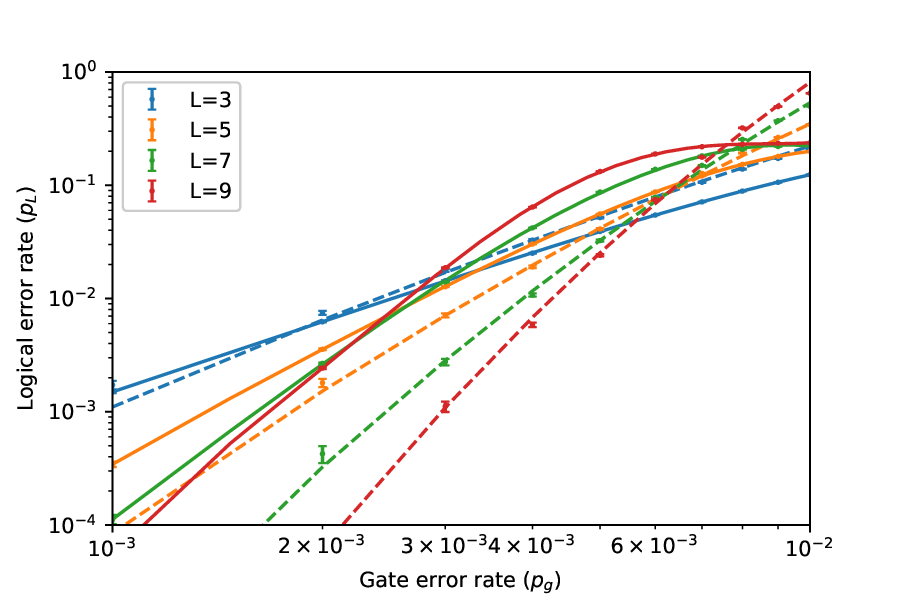}
    \caption{The logical error rate for single-shot checks with a single round of measurement (solid lines) and local checks with multiple rounds of measurement (dashed lines). The logical error rate $p_L$ for local checks is determined by decoding from $d$ rounds of noisy measurements and a last round of noiseless measurements. There is an apparent threshold of this model at 0.6\%. The logical error rate per round $p_L$ for the single-shot checks is estimated using the average from $N$ rounds,   $p_L=(1-(1-p_L(10))^{(1/10)})$. There is no threshold in this case and the optimal code depends on error rate.}
      \label{fig:ZX_comp}
\end{figure}

At the circuit level, we use a simplified $ZX$ error model for gates, a measurement error, and a depolarizing error between rounds. This model is motivated by gate errors that commute with the gate which can be dominant due to control errors \cite{Debroy_2020}. The standard circuit error model is outside of our current analysis due to the complexity that comes with error propagation \cite{dennis2002topological} and the need to schedule two-qubit gates.  

We start with a perfectly encoded state. Before each round of single-shot error correction, we assume each data qubit and syndrome qubit undergoes an independent depolarizing channel $C(\rho)=(1-p_g)\rho+\frac{1}{3}p_gX\rho X+\frac{1}{3}p_gY\rho Y+\frac{1}{3}p_gZ\rho Z$. We further assume a correlated $ZX$ error occurs every time a CNOT gate is applied with a probability $p_g$ where a $Z$ error occurs on the control qubit and an $X$ error occurs on the target qubit. We call $p_g$ the gate error rate. In this case, errors on syndrome qubits do not propagate to the data qubits \cite{dennis2002topological,PhysRevA.96.032341}. During the measurement, we assume idling qubits experience no error. After all the CNOT gates are applied, we attach measurement errors by independently flipping the measurement outcome of each syndrome qubit with probability $2p_g/3$. The primary purpose of the model is to increase syndrome measurement error with stabilizer weight. The lack of error propagation simplifies the fault-tolerance analysis but the application of a data qubit and measurement qubit error with probability $p_g$ is stronger than the standard two-qubit depolarizing model. This circuit model is inspired by trapped-ion qubits where qubits have long coherence times \cite{wang2017single,wang2021single} and gate errors are often systematic \cite{Debroy_2020}.

The phenomenological error models only considered $X$ errors on the qubits and only considered the detection and correction of those errors.  Here we need to detect and correct both types of errors to determine if a logical error occurs after the correction.

To compare with local check performance, we further calculate $p_L$ for both cases. Fig.~\ref{fig:ZX_comp} compares the local checks with multiple measurements and the single-shot checks with single measurements. The conventional method of local checks leads to an observed threshold at $p_g=$ 0.6\% for this model. The threshold is lower than the standard depolarizing error model since the two-qubit gate error is always correlated. The single-shot checks do not result in a threshold.  The optimal encoding level $L$ depends on the error $p_g$. For a given application, there remains a tradeoff between error correction cycle time and logical gate fidelity for these two approaches.

In this work, we investigated a method to systematically generate single-shot parity checks for topological codes based on Gaussian elimination. 
The main idea is to guarantee that each check can be flipped by a single-qubit error while the same error does not affect any other check. In phenomenological noise model simulations, our checks show single-shot error correction properties with a sustainable threshold at approximately 5.62\%, which is almost twice as high as that of the toric code at 2.9\% with repeated measurements [\onlinecite{wang2003confinement}]. 
Our simulation results also confirm that the local checks for toric codes shows no single-shot threshold behavior. When considering $ZX$ gate errors, depolarizing data errors, and measurement errors, the single-shot approach no longer has a threshold but instead an optimal code size for a given physical error. This leads to a tradeoff between logical qubit fidelity and the latency of quantum error correction. 

Single-shot checks for the surface code requires greater qubit connectivity than local checks and provides ample opportunities for error propagation using a standard depolarizing circuit-level error model. Error propagation can be controlled through encoded syndrome qubit such as cat states \cite{548464,PhysRevLett.77.3260} or through flag qubits\cite{PRXQuantum.1.010302}.   So far, there have been only a few studies on circuit-level simulation of single-shot error correction \cite{tremblay2022constant}. An obstacle for circuit-level simulation is the errors occurring between the measurements of two stabilizers in the same round of single-shot error correction. This leads to inconsistency of the syndrome information \cite{delfosse2021beyond} and more work needs to be performed with circuit-level errors to truly evaluate single-shot error correction.

The authors thank T. Tansuwannont, B. Pato, A. Nemec, M. Kang, M. Gutierrez, J. Campos, and E. Guttentag for the valuable discussions. This work was supported by ARO/LPS QCISS program (W911NF20S0004) and the NSF QLCI for Robust Quantum Simulation (353000072).

\appendix

\section{\label{sec:appenN1} Comparison of $N=1$ single-shot performance for local and single-shot checks}
We investigate the case of $N=1$ for both the local checks and the single-shot checks. As described in the main text we use MWPM for the qubit decoding. When studying the local checks, we also use the parity of syndrome to perform syndrome decoding. If the parity is even, we input noisy syndrome directly into the decoder for 2-D toric codes. If the parity is odd, we randomly flip one of the syndrome bits before passing it to qubit decoding. This is because an even syndrome parity is necessary for the qubit decoding on a toric code model due to the check redundancy\cite{higgott2022pymatching,kitaev2003fault,dennis2002topological}.

\begin{figure}

    \centering
    \includegraphics[width=0.45\textwidth]{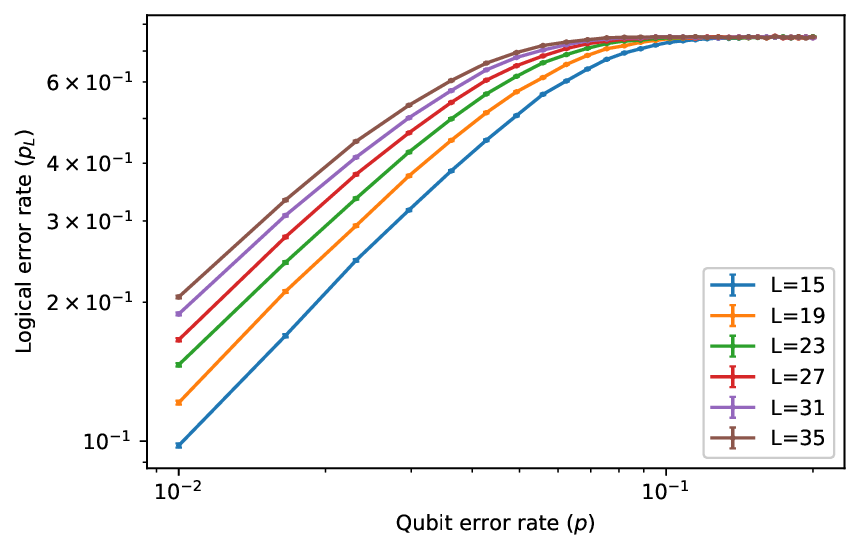}
    \caption{Local checks with one round of single-shot error correction ($N$=1) for code sizes $L$ from 15 to 35. No threshold behavior can be found as smaller code size shows stronger resilience of logical errors under all physical error rates. }
    \label{fig:toric}
\end{figure}

For the local checks, we collect the data from 80000 samples for local check with lattice sizes ranging from 15 to 35. As is seen in Fig.~\ref{fig:toric}, no threshold behavior occurs. For any qubit error rate, increasing code sizes always increases the corresponding logical error rate. This result confirms the statement that local check matrices have no single-shot properties [\onlinecite{campbell2019theory}]. 

\begin{figure}

    \centering
    \includegraphics[width=0.45\textwidth]{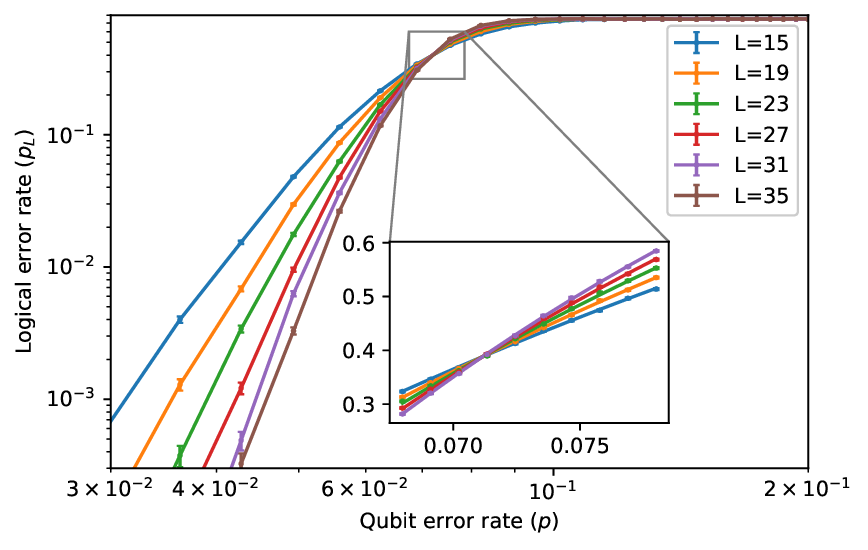}
    \caption{Single-shot checks with one round of single-shot error correction ($N$=1) for code sizes $L$ from 15 to 35. A clear threshold behavior can be seen. The error bars show the standard error of the mean values. The intersection occurs at the physical error rate 7.12\%. Below this probability, codes with larger code sizes have less logical error. Above this probability, codes with smaller code sizes outperform those with larger ones. The inset is the area close to the threshold. We follow \cite{PRXQuantum.2.020340} and fit the data around the threshold by $a_0+a_1x+a_2x^2$ where $x=(p-p_{\textrm{th}}(1))L^{1/\mu}$. Thus, we obtain the fitting values $a_0=0.388$, $a_1=3.280$, $a_2=-4.996$, $\mu=1.505$ and $p_{\textrm{th}}(1)=0.07116$. The results are obtained from 80000 samples. }
    \label{fig:N1}
\end{figure}

For $N=1$, we perform the Monte Carlo simulation for code lattice sizes ranging from $L=15$ to $L=35$ where each data point averages over up to 200000 samples for the single-shot checks. 
The results are presented in Fig.~\ref{fig:N1}. In contrast to the local code results in Fig.~\ref{fig:toric}, we can see that under a critical physical error probability, larger code sizes show lower logical error rates, which indicates errors are suppressed even though faulty syndrome are involved in the decoding process in this case. The threshold $p_{\textrm{th}}(1)$ is $7.12\%$.

\section{\label{sec:appen}Sustainable threshold for planar codes and rotated surface codes\protect}
\begin{figure*}
    \centering
    \subfloat[]{\includegraphics[width=0.45\textwidth]{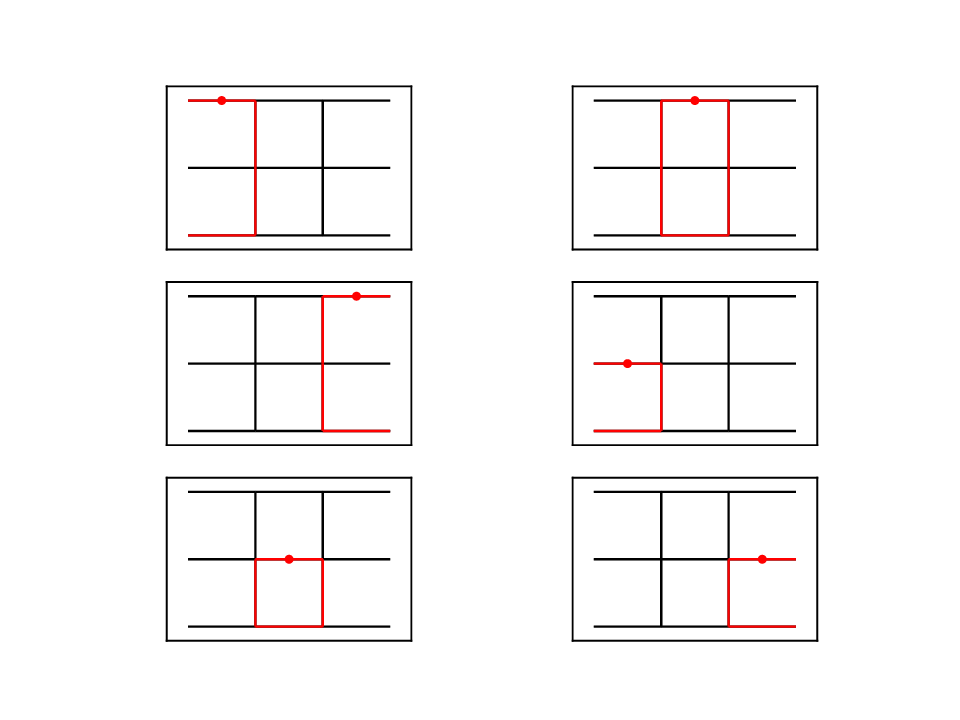}}\subfloat[]{\includegraphics[width=0.45\textwidth]{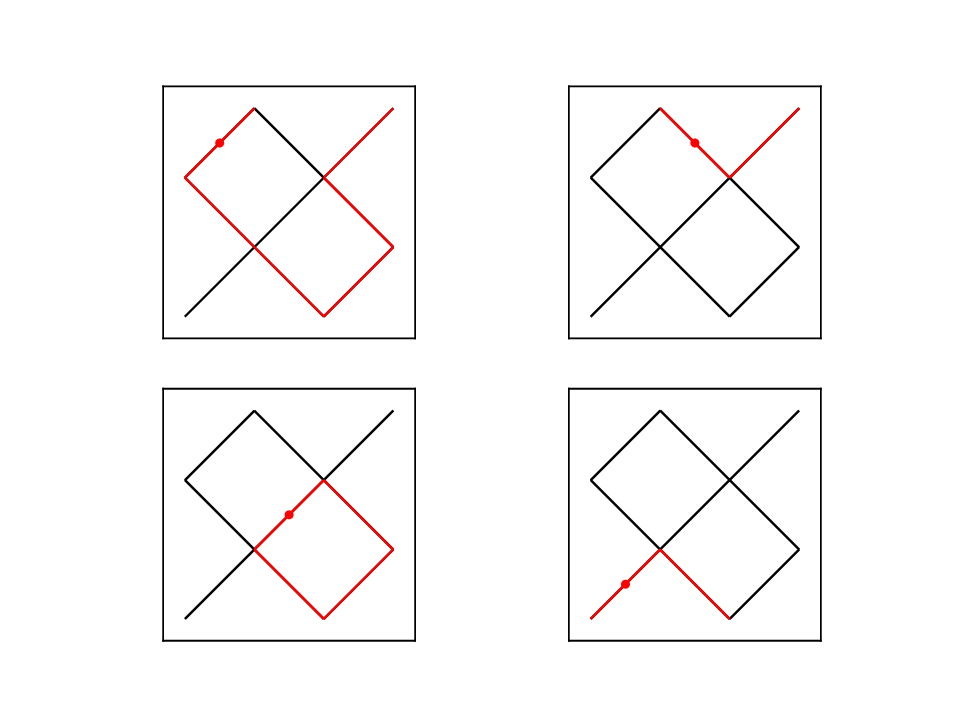}}
    \caption{The single-shot $Z$ checks for (a) planar codes and (b) rotated surface codes with distance $L=3$, obtained from Gaussian elimination of the parity check matrix. The qubits reside on the links. The red links are qubit acted on by each single-shot check. We mark the corresponding single-qubit $X$ error, $E_C$, with a red point on each graph.  }
    \label{fig:planar_Stab}
\end{figure*}

\begin{figure}
    \centering
    \subfloat[]{\includegraphics[width=0.45\textwidth]{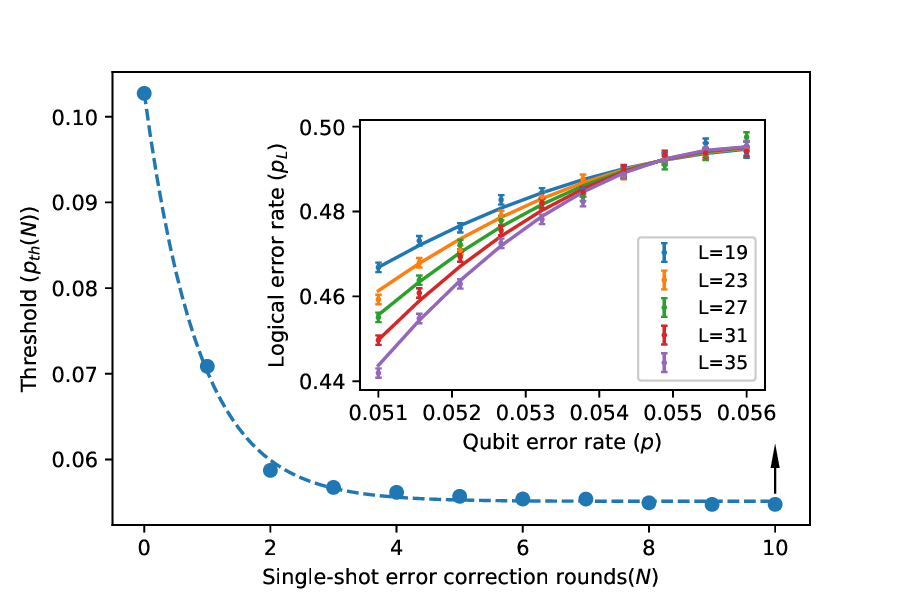}}
    \caption{Sustainable threshold for planar codes at $5.51\%$.}
    \label{fig:two_sus}
\end{figure}

\begin{figure}
    \centering
    \subfloat[]{\includegraphics[width=0.45\textwidth]{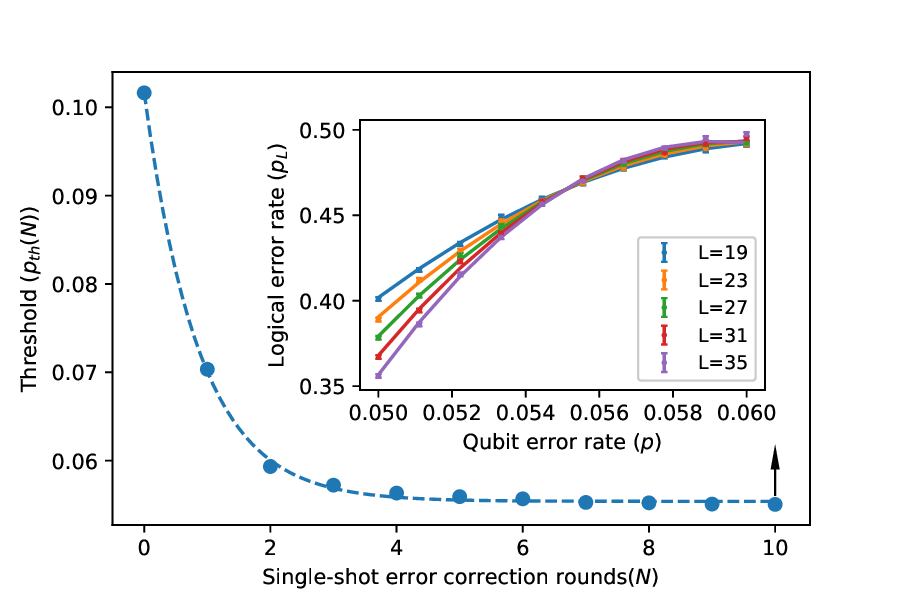}}
    \caption{Sustainable threshold for rotated surface codes at $5.54\%$.}
    \label{fig:sus_rotate}
\end{figure}

Local interaction is desirable in experiments, especially for superconducting qubits. Planar codes and rotated surface codes are adapted from toric codes by inserting boundary conditions that allow us to arrange the qubits in a plane. These codes emerged as a promising candidate for near-term physical implementations \cite{PhysRevA.90.062320,google2023suppressing,krinner2022realizing,PhysRevLett.129.030501}. To relate to experiments, we further perform single-shot error correction simulation on planar codes and rotated surface codes with single-shot checks found by Gaussian elimination. The single-shot stabilizers for both cases on $L=3$ lattices are given in Fig.~\ref{fig:planar_Stab} (a) and (b) respectively. We generate 200000 samples for each physical error rate with distances ranging from $L=19$ to $L=35$. The resulting sustainable thresholds are $5.51\%$ (Fig.~\ref{fig:two_sus}) and $5.54\%$ (Fig.~\ref{fig:sus_rotate}) respectively. These values are slightly lower than the toric code result, which is likely due to the different single-shot checks for these codes.

\FloatBarrier
\nocite{*}
\bibliography{apssamp}

\end{document}